# Comment on Consensus formation on a simplicial complex of opinions
# [Volume 397, 1 March 2014, Pages 111–120]

Vesna Berec[1]

[1]*Institute of Nuclear Sciences Vinca, University of Belgrade P.O. Box 522, Serbia*

*E-mail*: [1]vberec@vinca.rs

## Abstract

In the commented paper, the authors declare an analogy with the quantum mechanical pure states established through the application of the high-dimensional combinatorial Laplacian considering that the simplicial complex is in the pure state when it is formed by collection of the pure states. In their work, besides giving a completely erroneous analogy for the pure quantum mechanical state, which contradicts to very basic postulates of quantum mechanic, the authors clearly fail to provide and explain mathematical formalism behind their claims. Their intention is to rigorously prove that the existence of opinion space in all considered cases consists of the pure quantum states using incorrect normalization constant which does not produce trace equal to one. In this comment we prove out that their claims are erroneous. In order to show incorrectness of Maletic & Rajkovic model and conclusions we calculated and examined connections between different combinatorial structures of q-dimensional simplicial complexes and their Laplacian spectra and analyzed resulting symmetric and positive semidefinite matrices in the context of the density matrix of a quantum mechanical system. Importance of presented research is to further develop accurate quantitative topological-based characterization method for convertibility and distillation of density matrix.

# Introduction

Density-matrix description of quantum states is central both to the foundation of quantum mechanics and to quantum statistics and computation [1, 2]. Its importance has been demonstrated in various applications from quantum dynamics and optimization of information systems to general topology and graph theory [3], quantum cryptography and error correction [4], i.e., in any field where we are not aware of exact details of the state of our system or ensemble of systems. The density-matrix description is also applicable when dealing with a composite system that holds different qualities of information about its component subsystems. Keeping in mind that the density-matrix representation allows a partial description of a quantum system, assuming its general probabilistic (intrinsic) nature, one imposes a natural question: Can the density-matrix description of system states serve as a straightforward measure for the complexity analysis, especially in the field of spectral graph theory [5] and further: does the opinion representation of combinatorial Laplacians of simplicial complexes provide a direct link (analogy) with a pure quantum state represented in terms of a density matrix? One of the main reasons to impose such questions is the existed open potential of certain quantum models when restricted to Laplacian matrices, which allows resolving underling physical characteristics of topological structures relating to the spectral graph theory.

In spite of its general applicability, the density matrix implemented on graphs have been introduced quite recently, Braunstein *et al.* [5], as a restricted class of states with the aim to improve combinatorial optimization and to explore spectral features of separable quantum states. The density matrix of a graph can be always written as a uniform mixture of pure density matrices of graphs [5] showing the essential feature of the mixed quantum state, $\rho$. On contrary, the statement that the density matrices of graphs are normalized combinatorial laplacians with trace one is neither sufficient (because degree of purity of density matrix cannot be related to probabilistic attribute of the density matrix, i.e., a mixed state density matrix can also possess trace equal to one but only the pure quantum state holds the property that all the elements of an ensemble are in the same state), nor correct because it implicitly fails to relate complete state density matrix to its direct analog - a statistical mixture of pure states which is not a projector, i.e., the eigenvalues of $\rho$ and $\rho^2$ are not the same resulting that $\rho^2 \neq \rho$.

The explanation of the former statement stems from the basic postulates of quantum mechanics and at the same time disproves any analogy of pure quantum states with opinion states defined through Maletic & Rajkovic model of combinatorial Laplacians of simplicial complexes [6].

Namely, similar erroneous arguments drawn for implementation of the density matrix of pure quantum state in the scope of combinatorial algebraic topology have been given in [6]. In particular, there is formulated that the similar analogy can be established between the quantum mechanical pure state and the pure geometrical states of the opinion space in the simplicial combinatorial laplacian matrix. The other objection related to [6] refers to unsupported claims for entanglement existence that are exposed purely descriptive without any proves nor solid bases using overlaps called simplicial complex in the suggested model called "simplicial complex of opinions". Although ref. [6] give well dimension drawing explanation on social relations like "corruption–unemployment–high income inequality, etc., one of the problems is that the authors do not express the specific distinguishing property of entanglement referring to a quantum state, in spite of the authors attempt to define the pure geometrical states of the opinion space in analogy with the quantum mechanical pure states by taking the $q$-Laplacian matrix as a density matrix at dimension $q$. Instead, they identify entanglement with overlapping and freely characterize a collection of opinions as entanglement of overlapping opinions and shared judgments, by accounting former as the pure quantum states. Moreover, authors in ref. [6] do not distinguish boundaries between product states and entanglement nor any kind of correlation merely because they do not associate any dimensionality of the Hilbert space with the tensor product spaces to a space enclosed by the faces of each subcomplex and the higher dimensional simplex.

In order to address the problems arising from the commented paper, we first introduce preliminary set-theoretic concepts related to simplicial complex of graph and $q$-th combinatorial Laplacian, introducing higher dimensional simplicial complexes. A more detailed treatment of the subject is discussed in [7].

## Simplicial complex and a *q*-th combinatorial Laplacian

An abstract simplicial *complex* $\Delta$ on a set $X$ is represented by a family of finite subsets of $X$, closed under deletion of elements. Abstract simplicial complex can be realized as a geometric object in $\mathbb{R}^n$ called a (geometric) simplicial complex, which is a convex set spanned by $n+1$ affine independent simplices of specific dimensions. In particular, a graph represents a simplicial complex of dimension 1, called simplex-1. The convex hull of a subset of simplices represents a face of a simplex.

For $q \geq -1$, the $q$-skeleton $\Delta^{(q)}$, $\Delta^{(q)} \subseteq \Delta$, is realized by deletion of all faces of dimension greater than $q$. According to the set theoretic notation of graph complexes, 0-dimensional faces are called vertices and 1-dimensional faces are referred as edges. A simplicial complex $K$ is a collection of simplices where any face of a simplex of $K$ is an element of $K$ and the intersection of two simplices in $K$ is a face of both simplices. Given a set of oriented simplices $[v_0, v_1, ..., v_n]$, defined on the vector space whose basis is a set of ordered elements for each $q > 0$, the linear combinations of the basis elements are called chains, denoted as $C_q(K)$. The sequence of chains over the $q$ th-simplex of $K$ is represented as a chain complex spanned by the boundary operators: $\partial_q : C_q(K) \to C_{q-1}(K)$, where $\partial_q [v_0, v_1, ..., v_n] = \sum_{j=0}^{n}(-1)^j [v_0, v_1, ..., \widehat{v}_j, ..., v_n]$, and $\widehat{v}_j$ denotes vertex which is omitted from the oriented simplex.

The $q$ th-combinatorial Laplacian of a finite oriented simplicial complex $K$, for integer $q \geq 0$, is defined by the linear operator $\Delta_q : C_q \to C_q$ [8] as $\Delta_q = \partial_{q+1} \circ \partial_{q+1}^* + \partial_q^* \circ \partial_q$, where the $q$ th-combinatorial Laplacian matrix of $K$, denoted $\mathcal{L}_q$, is the matrix representation: $\mathcal{L}_q = \mathcal{B}_{q+1}\mathcal{B}_{q+1}^{\mathrm{T}} + \mathcal{B}_q^{\mathrm{T}}\mathcal{B}_q$ of operator $\Delta_q$, and $\mathcal{B}_q$, $\mathcal{B}_{q+1}$ are matrices of dimension $q$ and $q+1$, respectively.

The $q$ th-combinatorial Laplacian of a finite oriented simplicial complex $K$, for convenience, can be represented by the sum $\mathcal{L}_q = \mathcal{L}_q^{\mathrm{UP}} + \mathcal{L}_q^{\mathrm{LW}}$ [9], where $\mathcal{L}_q^{\mathrm{UP}} = \mathcal{B}_{q+1}\mathcal{B}_{q+1}^{\mathrm{T}}$ and $\mathcal{L}_q^{\mathrm{LW}} = \mathcal{B}_q^{\mathrm{T}}\mathcal{B}_q$, index upper and lower degree for a $q$-simplex, respectively. Number of rows and columns in matrix $\mathcal{B}_q$ corresponds to the number of $(q-1)$-simplices and the number of $q$-simplices in $K$, respectively.

According to the Laplacian Matrix Theorem [10] if $K$ is a finite oriented simplicial complex, where $q$ is an integer defined for $0 < q \leq \dim K$, where $\dim K$ is the dimension of simplicial complex and $[\sigma_1, \sigma_2, ..., \sigma_n]$ are the $q$-simplices of $K$ for $i, j \in \{1, 2, ..., n\}$, then the $q$ th-combinatorial Laplacian matrix of $K$, $\mathcal{L}_q = \mathcal{B}_{q+1}\mathcal{B}_{q+1}^{\mathrm{T}} + \mathcal{B}_q^{\mathrm{T}}\mathcal{B}_q$, have the following entries:

$$q = 0, \, (\mathcal{L}_q)_{ij} = \begin{cases} \deg_U(\sigma_i) = \deg_G(v_i), & \text{if } i = j, \\ -1, & \text{if } \sigma_i, \sigma_j \text{ are upper adjacent}, \\ 0, & \text{otherwise}. \end{cases} \quad (1)$$

$$q > 0, \, (\mathcal{L}_q)_{i,j} = \begin{cases} \deg_U(\sigma_i) + q + 1, & \text{if } i = j, \\ 1, & \text{if } i \neq j, \text{ and } \sigma_i, \sigma_j \text{ are not upper adjacent} \\ & \text{but have a similar common lower simplex}, \\ -1, & \text{if } i \neq j, \text{ and } \sigma_i, \sigma_j \text{ are not upper adjacent} \\ & \text{but have a disimilar common lower simplex}, \\ 0, & \text{if } i \neq j, \text{ and } \sigma_i, \sigma_j \text{ are uper adjacent} \\ & \text{or are not lower adjacent}. \end{cases}$$

# Combinatorial Laplacian matrix of a graph

**Definition.** A (simple) graph $G(V,E)$ consists of a finite set $V(G)$ of vertices and a collection of size-2 subsets of $V$ called edges, $E \subseteq \binom{V}{2}$. A graph is refereed to as simple if it contains no loops. The number of edges incident on a vertex $v_i \in V(G)$ represents the degree of a ith vertex, denoted $\deg_G(v_i)$. An *isolated* vertex is of degree 0.

Let $G$ be a finite graph with $V(G) = \{v_1, v_2, ..., v_n\}$, and the edge set $E$ containing all pairs $(v_i, v_j)$ of vertices in $V$. The combinatorial Laplacian matrix of a graph $G(V,E)$ is the $n \times n$ matrix denoted as $L_G$ (also referred to as Laplacian) represented by the following elements:

$$L_{ij} = \begin{cases} \deg_G(v_i), & \text{if } i = j, \\ -1, & \text{if } v_i \sim v_j \ (i \text{ adjacent to } j), \\ 0, & \text{otherwise,} \end{cases} \qquad (2)$$

for all $i, j \in \{1, 2, ..., n\}$, where $\deg_G(v_i)$ is the degree of the ith vertex, showing the number of edges in $G$ containing $v_i$. From the aspect of relations of adjacency, a combinatorial Laplacian matrix of a graph can be written as $L_G = D - \mathcal{A}$, where $D$ is the diagonal matrix with $i = j$ entries, where $\deg_G(v_i) = d_i$, and $\mathcal{A}$ is the adjacency matrix.

The entries of adjacency matrix $\mathcal{A}$ of a graph $G(V,E)$, are specified by the vertices of $G$, as:

$$\mathcal{A}(v_i, v_j) = \begin{cases} 1, & \text{if } v_i, v_j \in E, \\ 0, & \text{otherwise,} \end{cases} \qquad (3)$$

for all $i, j \in \{1, 2, ..., n\}$.

When the graph is oriented and represented as a 1-dimensional simplicial complex (1-simplex), the combinatorial Laplacian matrix of a graph $L_G$ corresponds to the matrix representation of the boundary operator homomorphism: $\partial_q = \partial_1 : C_1(K) \to C_0(K)$ defined for the corresponding chain groups $C_q(K)$ of the simplicial complex $K$ [7], associated with that graph, which is a 0-th combinatorial Laplacian matrix $\mathcal{L}_0 = \mathcal{B}_1 \mathcal{B}_1^T$ [7, 9]. $\mathcal{B}_1^T$ is the matrix representation of the adjoint boundary operator, $\partial_1^*$, with respect to this same ordered basis. Number of rows and columns in matrix $\mathcal{B}_1$ corresponds to the number of $0$-simplices and to the number of 1-simplices in $K$, respectively. For $q=0$, and $K$ without singletons, the boundary operator $\partial_q : C_q(K) \to C_{q-1}(K)$ defines the zero map.

## Results

Let us inspect the normalized combinatorial Laplacian of simplicial complex as defined in [6].

*Example 1.*

In particular, by considering the directed simple graph $G(V,E)$ whose vertex set and edge set are given respectively by $V_G = \{1,2\}$, and $E_G = \{\{1,2\}\}$, the corresponding Laplacian matrix $L_G = D - \mathcal{A}$ of a graph $G$ is equal to a 0-th combinatorial Laplacian matrix of a 1-simplex, whose elements are given by Eq. (1):

$$(\mathcal{L}_0)_{ij} = \begin{cases} \deg_G(v_i), & \text{if } i = j, \\ -1, & \text{if } v_i, v_j \text{ are upper adjacent}, \\ 0, & \text{otherwise}, \end{cases} \quad (4)$$

for all $i,j \in \{1,2\}$.

Precisely, the representation of $q^{th}$ Laplacian matrix of simplicial complex $K$ [6] which is in the above case 1-simplex (since finite simple graphs can be seen as simplicial complexes of dimension 1, where one link connects two nodes of graph) is given by Eq. (1), as

$$\mathcal{L}_0(K) = \mathcal{B}_1 \mathcal{B}_1^T = \begin{pmatrix} 1 & -1 \\ -1 & 1 \end{pmatrix} \Leftrightarrow L_G = D - \mathcal{A} = \begin{pmatrix} 1 & 0 \\ 0 & 1 \end{pmatrix} - \begin{pmatrix} 0 & 1 \\ 1 & 0 \end{pmatrix} = \begin{pmatrix} 1 & -1 \\ -1 & 1 \end{pmatrix}, \quad (5)$$

where matrix $\mathcal{L}_0$ of a simplicial complex $K$ clearly corresponds to a $L_G = D - \mathcal{A}$ (0-th combinatorial laplacian matrix of $K$ equal to a combinatorial laplacian matrix of a graph), i.e., $\mathcal{B}_1$ equals to incidence matrix of $G$. Furthermore, the laplacian of a graph, $L_G$, can be considered as a density matrix only if it is a positive semidefinite, trace-one, hermitian matrix [5]. That means that the density matrix can be associated to a 0-th combinatorial laplacian matrix of a simplicial complex (equivalent to a laplacian matrix of a graph) if latter is normalized by the sum of its diagonal elements, as it is already defined by Braunstein *et al*. [5].

Without loss of generality only a $q=1$ -dimensional simplicial complex (1-simplex) as a positive semidefinite hermitian matrix can be associated to a density matrix, but *if and only if* it is properly scaled by the degree-sum of $G$ [5], which is in fact equivalent to the normalization constant $\frac{1}{\text{Tr}(L_G)}$.

Without taking into account that graph is the 1-dimensional simplicial complex $(q=1)$, which zero dimensional Laplacian matrix $\mathcal{L}_0(K)$ clearly corresponds to a combinatorial laplacian matrix $L_G$ of graph, authors in [6] firstly, abusing this notation, identify a "$q$-Laplacian matrix" ($L_q$) as a density matrix at an arbitrary dimension $q$. Note that dimension $q$ is associated to a simplicial complex (of opinions). Secondly, authors, in their definition of a $q$-Laplacian matrix as a density matrix at dimension $q$, choose incorrect normalization constant which does not provide trace to one

$$L_q^n = A \cdot L_q, \quad A = \frac{\sum_i n_i d_i}{\sum_i n_i d_i^2}, \tag{6}$$

where $d_i$ are diagonal elements and $n_i$ is multiplicity of the $d_i$-th diagonal element of the $q$-Laplacian matrix $L_q$, respectively; $A$ is the normalization constant, and according to authors claims [6] $L_q^n$ is the "properly" normalized Laplacian matrix for dimension $q$.

Authors than freely claim that a simplicial complex of opinions (representing a density matrix) is formed by the pure states if the following relation is satisfied:

$$\text{Tr}\left(L_q^n\right) = \text{Tr}\left(\left[L_q^n\right]^2\right), \tag{7}$$

i.e., in this case authors say that the simplicial complex is in the pure state if it is formed by the collection of disconnected simplices, disregarding the fact that the boundary operator of dimension zero, $\partial_0$, in such case defines an augmentation $\partial_0 = \varepsilon : \tilde{C}_0(K) \to \tilde{C}_{-1}(K) = \mathbb{Z}$ [7] defined by $\varepsilon([v]) = 1$ for every vertex $v$ in simplicial complex $K$, producing a mapping $\sum_i n_i \{v_i\} \to \sum_i n_i$ which is not associable to a density matrix. In fact, defining $K$ to be a simplicial complex, there exists a boundary operator $\partial_q$ which determines homomorphisms $\partial_q : C_q(K) \to C_{q-1}(K)$ for each integer $q$ between the chain groups of $K$, where $C_q(K)$ is the qth chain group of the simplicial complex $K$ such that $C_q(K) = \{\emptyset\}$ when $q < 0$ or $q > \dim K$ (where $\dim K$ denotes the dimension of the simplicial complex $K$). For a general case of $0 < q \leq \dim K$, if $\sigma$ is an oriented $q$-simplex of $K$ then $\partial_q(\sigma)$ is a $(q-1)$-chain that represents a sum of the $(q-1)$-faces of $\sigma$, where orientation of each face is determined by the orientation of $\sigma$. Thus, $\partial_q = 0$ if $q \leq 0$ or $q > \dim K$.

In case of a $q$-dimensional simplicial complex $K$ which includes singletons forming a 0-dimensional simplicial complex (0-simplex can be identified with a set of points (called 0-simplices, vertices or nodes) or simply with a point in space (node)), the augmented oriented chain complex [7] of $K$ over $f$-vector space represented by the $q$-th chain group of the simplicial complex $\tilde{C}_q(K)$ is given by:

$$0 \to f^{\tilde{C}_{n-1}(K)} \xrightarrow{\partial_{n-1}} \ldots \to f^{\tilde{C}_i(K)} \xrightarrow{\partial_i} f^{\tilde{C}_{i-1}(K)} \to \ldots \to f^{\tilde{C}_0(K)} \xrightarrow{\partial_0 = \varepsilon} f^{\tilde{C}_{-1}(K) = \mathbb{Z}} \to 0, \quad q = 0, 1, \ldots, n-1,$$

where $\sigma \in \tilde{C}_q(K)$ and $\partial_q(e_\sigma) := \sum_{j \in \sigma} \text{sign}(j, \sigma) \, e_{\sigma \setminus j}$, where $e_\sigma$ denotes the corresponding basis in the $f$-vector space, $f^{\tilde{C}_q(K)}$. In this case only if $q > n-1$ or $q < -1$, then $f^{\tilde{C}_q(K)} := 0$ and $\partial_q := 0$. As a result, the singletons from the 0-simplicial complex, as a part of the $q$-dimensional simplicial complex $K$, are associated to a matrix representation $\mathcal{L}_{q=0}^{\text{LW}} = \mathcal{B}_0^T \mathcal{B}_0$ [7, 9], where $\mathcal{B}_0$ is a row vector of $n$ elements (all entries are "1"). Due to its trace $\text{Tr}(\cdot) > 1$ obtained $n \times n$ matrix is not associable to a density matrix, nor can it be properly normalized to give the trace one according to the normalization constant from [6]. Furthermore, the augmentation: $\varepsilon \partial_1 = 0$ of a chain complex $\tilde{C}_q(K)$ over $\mathbb{Z}$ defines epimorphism $\varepsilon : \tilde{C}_0(K) \to \tilde{C}_{-1}(K) = \mathbb{Z}$, which spans $n_i$ number of vertices $v_i$ in complex $K$ into the sum $\sum_i n_i \{v_i\} \to \sum_i n_i$, which does not represent a density matrix. Note that $K$ is 0-connected if and only if it is path-connected, i.e., it represents a 1-simplex [3, 7].

Now, let us pay attention to the normalized Laplacian matrix described in [6].

Setting the normalization constant $A$ of the above $q$-Laplacian matrix, as defined in Maletic & Rajkovic model [6]:

$$A = \frac{\sum_i n_i d_i}{\sum_i n_i d_i^2},  \qquad (8)$$

where $d_i$ are diagonal elements of matrix $L_q$, and $n_i$ is multiplicity of the $d_i$ –th diagonal element (the multiplicities $n_i(a_{ii})$ are the number of times in which the diagonal element $d_i$ appears in the diagonal of the matrix $L_q$), fails to provide normalization to trace 1, with direct repercussion that normalized $q$-Laplacian matrix from [6] cannot be considered as a density matrix, and certainly cannot represent a pure quantum state.

*Proof:*

Let us first determine the $q$-Laplacian matrix according to [6], which is presented in example 1. Since a graph *G* can be viewed as a simplicial complex of dimension 1, the matrix $\mathcal{L}_0$ is the same as the zero-dimensional Laplacian matrix of *G* as a simplicial complex: $\mathcal{L}_0(K) = L_{q=0}$.

Now we shall determine the "normalized Laplacian matrix for dimension *q*" by applying the normalization constant as it is stated in [6]:

$$L^n_{q=0} = A \cdot L_{q=0} = \frac{\sum_i n_i d_i}{\sum_i n_i d_i^2} \cdot \begin{pmatrix} 1 & -1 \\ -1 & 1 \end{pmatrix} = 1 \cdot \begin{pmatrix} 1 & -1 \\ -1 & 1 \end{pmatrix}, \qquad (9)$$

where matrix diagonal elements $d_i = d_{a_{ii}}$, and their multiplicities $n_i(a_{ii})$, of the matrix $L_q$, are:

$$d_1 = 1, \; n_1 = 2; \; d_2 = 1, \; n_2 = 2; \;\text{resulting that}\; L^n_{q=0} = 1 \cdot \begin{pmatrix} 1 & -1 \\ -1 & 1 \end{pmatrix} \Rightarrow \text{Tr}(A \cdot L_{q=0}) = 2.$$

Above normalization according to [6] instead of tracing a matrix to one in fact normalizes only a constant $A$ to one. Direct repercussion is that above trivial example of the Laplacian matrix of a graph ($q$-Laplacian matrix $L^n_{q=0}$) after normalization model, suggested by Maletic & Rajkovic [6], is applied cannot be considered as a density matrix [5] at all, nor as a pure quantum state density matrix, which absolutely disqualifies considered model. Final proof that the "trace" condition is not satisfied and that no analogy between $L^n_{q=0}$ and the pure quantum state density matrix can exist is the test of eigenvalues of the $q$-Laplacian matrix from Eq. (9), defined and normalized according to [6]:

$$L^n_{q=0} = \begin{pmatrix} 1 & -1 \\ -1 & 1 \end{pmatrix} \text{ where } \begin{cases} \text{eigenvalue } \lambda = 0 \text{ (algebraic multiplicity } m_a = 1), \\ \text{eigenvalue } \lambda = 2 \text{ (algebraic multiplicity } m_a = 1). \end{cases}$$

From the fact that the pure quantum state is a projector and can have only one non-zero eigenvalue (i.e., of algebraic multiplicity one) which is thus strictly equal to 1 while all the others are zero, it is clear that normalized Laplacian matrix $L^n_{q=0}$ with its eigenvalues $\lambda = 0$ and $\lambda = 2$, considering Maletic & Rajkovic model [6], is not a pure quantum state. □

*Example 2.*

Let $K$ be the oriented simplicial complex given in figure 1, then the matrix representation of the $q^{th}$ Laplacian matrix of simplicial complex $K$ is $\mathcal{L}_1(K)$:

$$\mathcal{L}_1(K) = \mathcal{B}_2 \mathcal{B}_2^T + \mathcal{B}_1^T \mathcal{B}_1 = \begin{pmatrix} 1 & -1 & 1 \\ -1 & 1 & -1 \\ 1 & -1 & 1 \end{pmatrix} + \begin{pmatrix} 2 & 1 & -1 \\ 1 & 2 & 1 \\ -1 & 1 & 2 \end{pmatrix} = \begin{pmatrix} 3 & 0 & 0 \\ 0 & 3 & 0 \\ 0 & 0 & 3 \end{pmatrix}. \quad (10)$$

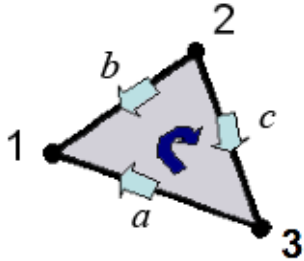

Figure 1. Representation of 2-simplex.

Again, the properly normalized Laplacian matrix for dimension $q$ [6], after the normalization is applied over diagonal elements $d_i = (d_1, d_2, d_3)$, where $d_1 = 3$, $d_2 = 3$, $d_3 = 3$; and their multiplicities on a diagonal $n_i = (n_1, n_2, n_3)$, where $n_1 = 3$, $n_2 = 3$, $n_3 = 3$; is given by

$$L_{q=1}^{n} = A \cdot L_{q=1} = \frac{\sum_{i} n_i d_i}{\sum_{i} n_i d_i^{2}} \cdot \begin{pmatrix} 3 & 0 & 0 \\ 0 & 3 & 0 \\ 0 & 0 & 3 \end{pmatrix} = \frac{1}{3} \begin{pmatrix} 3 & 0 & 0 \\ 0 & 3 & 0 \\ 0 & 0 & 3 \end{pmatrix}, \tag{11}$$

where eigenvalues of $L_{q=1}^{n}$ matrix with corresponding algebraic multiplicities are $\lambda = 1$, $m_a = 3$, respectively.

Note that obtained $L_{q=1}$ and $L_{q=1}^{n}$ matrix in given case are diagonal matrices. Determinant of a diagonal matrix is a product of main diagonal elements. Hence, for $L_{q=1}^{n}$

$$\det\left(L_{q=1}^{n} - \lambda I\right) = \begin{pmatrix} 1-\lambda & 0 & 0 \\ 0 & 1-\lambda & 0 \\ 0 & 0 & 1-\lambda \end{pmatrix} = (1-\lambda)(1-\lambda)(1-\lambda) = 0, \tag{12}$$

$\Rightarrow \lambda_1 = 1, \lambda_2 = 1, \lambda_3 = 1; \; m_a = 3.$

The resulting normalized matrix $L_{q=1}^{n}$ does not satisfy conditions for a density matrix or the pure quantum mechanical state. We get the trace value $\mathrm{Tr}(\cdot) = 3$, i.e., $\mathrm{Tr}\left(L_{q=1}^{n} = A \cdot L_{q=1}\right) \neq 1$, which automatically disproves analogy of presented simplicial complex of opinions and normalized $q^{\mathrm{th}}$ Laplacian matrix with the density matrix and the pure quantum mechanical state, claimed in [6]. Again, stronger proof that after normalization [6] obtained matrix (Eq. 12) cannot represent the pure quantum state, is provided by inspection of its eigenvalues, which correspond to elements on diagonal, $d_i$, (their algebraic multiplicities are equal to the number of times each $d_i$ appear, recall the property of diagonal symmetric matrix), where eigenvalues $\lambda = 1$ with algebraic multiplicity $m_a = 3$ do not meet criteria for the pure quantum state density matrix. □

Note, that result of the trace is same ($\mathrm{Tr}(\cdot) = 3$) if we apply normalization constant $A$ (Eqs. (6, 8)) only to a matrix representation $\mathcal{L}_1^{\mathrm{LW}}(K) = \mathcal{B}_1^{T} \mathcal{B}_1$ associated to $\partial_1$ boundary operator, of example 2. (Eq. 10), where matrix diagonal elements $d_i = \left(d_{a_{11}}, d_{a_{22}}\right)$ and their multiplicities $n_i = \left(n_{a_{11}}, n_{a_{22}}\right)$ are: $d_1 = 2, n_1 = 3; \; d_2 = 2, n_2 = 3; \; d_3 = 2, n_3 = 3$. In that case, after the normalization [6], eigenvalues of obtained matrix are: $\lambda_1 = 3, \lambda_2 = 3$ of algebraic multiplicity $m_a = 2$ and $\lambda = 0$ of algebraic multiplicity $m_a = 1$, disproving again applied model presented in [6], which does not meet criteria for a density matrix and clearly does not provide analogy with the pure quantum mechanical state.

## Discussion

The authors declare that opinion space in all considered cases consists of pure states at the end of their simulation. Conversely, simple analysis of the opinion space (mapped to simplicial complex) shows oppositely that the sum over all diagonal elements of which consensus state is formed, implying that simplicial complex of opinions is geometrically connected, results in trace: $\text{Tr} > 1$. Namely, parameter $S_{max}$ referring to a maximal number of different realized opinions associated to agents during the simulation, coincide with (normalized) diagonal matrix elements. In this case: between 0.3 and 1 (figure 5. in [6]). We shall further impose known constraints concerning the trace over density matrix in order to explain and adequately estimate incorrectness of the model presented in [6].

**1.** Recall that the mixed state density matrix $\rho$ always holds property $\text{Tr}(\rho)^2 < 1$. This means that quantitatively neither opinion space is a mixture of pure states, nor simplicial complex of opinions is a pure state taken by suggested model. The core of the problem is that authors fallaciously apply relation (7) to test whether a matrix which does not represent a density matrix at all (inappropriate normalization constant $A$ always produces trace unequal to one $\text{Tr}(\cdot) \neq 1$), describes a system (simplicial complex) formed by pure states in analogy to the quantum mechanical pure states. Recall, authors in the commented paper account the $q$-Laplacian matrix as a density matrix at dimension $q$, when they claim that the simplicial complex of opinions is formed by the pure states. Thus, they apply the "trace" condition on normalized $q$-Laplacian matrices which can not have the sum of eigenvalues equal to one after performed normalization [6], i.e., without taking into account that for the pure quantum state sum of its eigenvalues must be equal to the trace (one).

**2.** Qualitatively, presented model imposes more serious flaws. A property of the pure quantum state density matrix is that contribution of all its diagonal elements must sum to one ($\text{Tr}(\rho) = 1$) because pure state is a projector onto a one-dimensional subspace of the Hilbert space with only one nonzero eigenvalue - equal to unity. In another words, the pure quantum state possesses only one non-zero eigenvalue (and strictly equal to 1) while all the other are zero, directly implying idempotency relation $\rho^2 = \rho$ where $\text{Tr}(\rho)^2 = 1$. Conversely, the properties of a mixed state are that $\text{Tr}(\rho)^2 < 1$ and it cannot be expressed in terms of one pure state only. Hence, a collection of disconnected pure states (of opinions) can only represent a mixed state, a statistical mixture in a probabilistic way, in contrast to claims raised in [6]. Moreover, such collection of disconnected simplices cannot be associated to a density matrix.

During the act of connecting among the opinions, when the specific bond (tie, communication) is realized between the simplices [7], their property of purity is lost causing the instantaneous transformation in the topology of opinion space (monitored by evolution of the Q-vector).

**3.** Recall that authors [6] in their claims even contradict to basic postulate of quantum mechanic saying that the simplicial complex (as a system) is in the pure state analog to quantum mechanical pure state, if it is formed by the collection of simplices, which are disconnected. Keeping in mind that disconnected simplices can be easily associated to a disconnected graph (1-simplex), that would mean that the simplicial complex is in the pure state when it is formed by the collection of many pure states, which is erroneous statement keeping in mind what is specific feature of the mixed quantum state. Then authors even contradict to themselves by associating different realized opinions to agents, where the simplicial complex of opinions is geometrically connected at all levels of connectivity, and by saying that such simplicial complex of opinions then consists of pure states. This is also not true because in presented case different construction applies to representations defined over an algebraically closed field of any characteristic implying more than one such "connections" to a specific single pure state (in consistency to example of a seemingly "paradoxical" situation described in commented article) thus, in that case recall that only a mixed state, again, can be associated with more than one state. The connections are those who initialize transformation between mixed states and change the state representation to a new structure.

At the end, the described "paradoxical" situation where the growing number of opinions causes increase of the number of agents (and the number of individuals) which adopt the same opinion, represent an expected reflection of system tendency to relax towards state of maximum likelihood information (where the limit corresponding to a maximum information would be an isolated single state, which cannot be represented by density matrix applying the model suggested in [6]).

## Summary

In a recent article [Volume 397, 1 March 2014, Pages 111–120] the authors have derived hierarchy relations between simplicial complexes of opinions and wrongly claimed analogy with the quantum mechanical pure states established through the application of the high-dimensional combinatorial Laplacian (considering the incorrect normalization constant which does not produce trace equal to one). Incorrectness of Maletic & Rajkovic model and conclusions are shown. We have prove out that their claims are erroneous and result in analytic counterexamples. We have calculated and examined connections between different combinatorial structures of $q$-dimensional simplicial complexes and their Laplacian spectra and analyzed resulting symmetric and positive semidefinite matrices in the context of the density matrix of a quantum mechanical system applicable to the $n$-dimensional Hilbert space. The aim of research is to incorporate deterministic aspect of algebraic transformation in quantum information theory and to develop accurate quantitative topological-based characterization method for convertibility and distillation of density matrix under scalable LOCC (local operations and classical communications) for efficient implementation of information processing schemes.

**Acknowledgment**

This work was supported by the Ministry of Education, Science and Technological Development of the Republic of Serbia.